\documentclass[]{spie}  

\usepackage[]{amsmath}
\usepackage[]{amssymb}
\usepackage[]{graphicx}

\title{Electronic structure and optical properties of metallic nanoshells}

\author{P. Nordlander and E. Prodan
\skiplinehalf
Department of Physics and  Rice Quantum Institute, M.S. 61, Rice
University\\
Houston, TX 77251-1892, USA\\
}

\authorinfo{Further author information: (Send correspondence to P.N.)\\P.N.:
E-mail: nordland@rice.edu, Telephone: 1 713 348 5171\\  E.P.: E-mail:
emprodan@rice.edu, Telephone: 1 713 348 3541}

\begin{document}
\renewcommand{\textheight}{21.7cm}
\maketitle

\begin{abstract}
The electronic structure and optical properties of metallic nanoshells are investigated
using  a jellium model and the Time Dependent Local Density Approximation (TDLDA). 
An efficient numerical implementation enables applications to nanoshells of realistic size 
with up to a million electrons. We demonstrate how a frequency dependent background
polarizability of the jellium shell can be included in the TDLDA formalism. The energies
of the plasmon resonances are calculated for nanoshells of different sizes and with
different dielectric cores, dielectric embedding media, and dielectric shell backgrounds.
The plasmon energies are found to be in good agreement with the results
from classical Mie scattering theory using a Drude dielectric function. 
A comparison with experimental data shows excellent agreement between theory and 
the measured frequency dependent absorption spectra.
\end{abstract}


\section{INTRODUCTION}
\label{sect:intro}

The optical properties of metallic nanoparticles are of considerable fundamental
and technological interest.\cite{Brongersma03NAT}
A particularly interesting nanoparticle is the metallic
nanoshell,\cite{AverittetAl97PRL}
which consists of a metallic layer grown over a solid dielectric core. The
plasmon frequency, which determines the nanoshell optical properties,
can be tuned
over a wide spectral range by simply varying the ratio of inner to outer
diameter of the shell. The tunability of the optical properties of
metallic nanoshells enables several important applications
such as resonant photooxidation inhibitors,\cite{HaleetAl01APL}
optical triggers for drug delivery
implants,\cite{SershenetAl01AP,SershenetAl00JBMR}
environmental sensors,\cite{SunXia02AC} and Raman
sensors.\cite{JacksonetAl03APL}

Although classical Mie scattering seems to account well for the optical
properties of metallic nanoshells, there is clearly need for a more
microscopic description. A microscopic understanding of the electronic properties
of nanoshells is a prerequisite for a systematic and rational modification of their
electronic and optical properties.
We have recently started the development of a fully quantum mechanical
model for the electronic properties of metallic nanoshell using density functional
theory.\cite{ProdanNordlander01CPL,ProdanNordlander02CPL,ProdanetAl02CPL,%
NordlanderProdan02SPIE,ProdanetAl03CPL,ProdanNordlander03NL}
In this paper we will review the general features of our approach
and present some new results. The emphasis will be on the effects of dielectric
backgrounds on the electronic structure and optical properties of metallic nanoshells. 

\section{Electronic Structure}

TDLDA is a well established method, \cite{ZangwillSoven80PRA} which  allows 
surprisingly accurate calculations of the electronic and optical properties of nanoparticles. 
The method was extensively applied to small metal clusters and proven to be in excellent agreement
with more exact calculations and experiment.\cite{Heer93RMP,Brack93RMP} The method was recently 
implemented to shell geometries and proven to be in good agreement with the existent classical and 
semi-classical 
approaches.\cite{ProdanetAl02CPL,NordlanderProdan02SPIE,ProdanetAl03CPL,ProdanNordlander03NL}
TDLDA involves two steps. The first step consists of self-consistent calculation of the electronic 
structure and the independent electron response functions of the nanoparticle. The second step 
consists of solving self-consistently the RPA equation for the screened response function of the system. 
Our implementation of the first step to shell geometries has been discussed in details
previously,\cite{ProdanNordlander02CPL,ProdanetAl02CPL,NordlanderProdan02SPIE} and therefore only the
main points will be presented here.

\begin{figure}[h]
\begin{center}
\begin{tabular}{c}
\includegraphics[height=10cm]{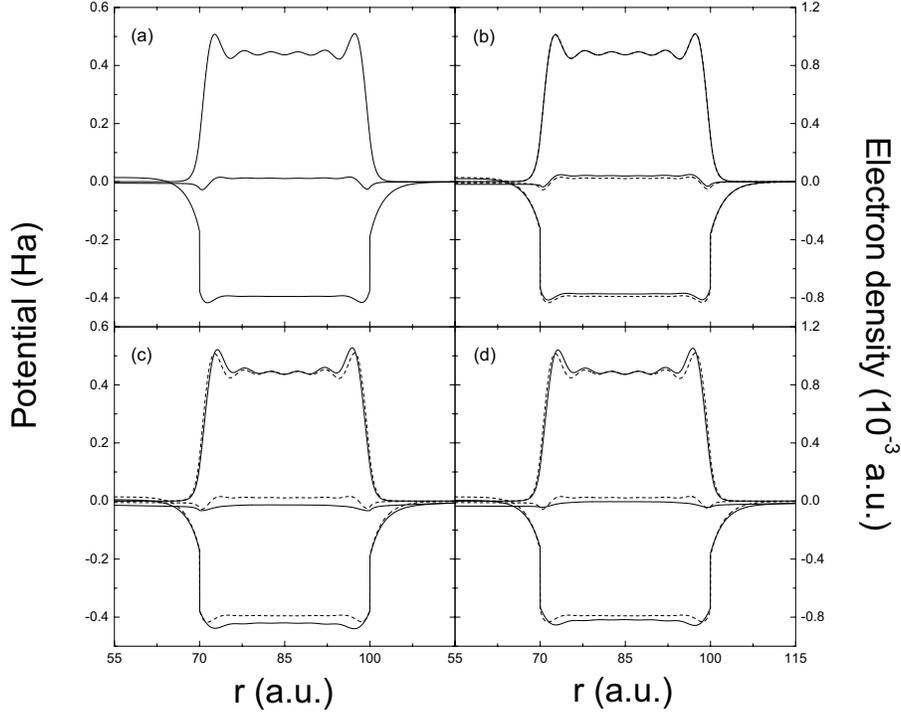}
\end{tabular}   
\end{center}
\caption[example]
{The effect of different dielectrics on the electronic structure of a (70,100)~a.u.
gold nanoshell. The curves in each panel represent, from the bottom and up, the   
effective potential, the Coulomb potential (the values are on the left axis) and
the electron density (the values on the right axis). Different panels
correspond to: a) $\varepsilon_{C}=1$, $\varepsilon_{E}=1$ and
$\varepsilon_{J}=1$; b) $\varepsilon_{C}=5$,
$\varepsilon_{E}=2$ and $\varepsilon_{J}=1$; c) $\varepsilon_{C}=1$, $\varepsilon_{E}=1$ and
$\varepsilon_{J}=8$; d) $\varepsilon_{C}=5$, $\varepsilon_{E}=2$ and $\varepsilon_{J}=8$.
For comparison, in panels b)-d) we have included with dashed lines the electronic
structure when no dielectrics are present, i.e. $\varepsilon_{C}=\varepsilon_{E}=\varepsilon_{J}=1$.
}
\end{figure}

The electronic structure of the metallic nanoshells is determined from the Kohn-Sham equations
\begin{equation}
\left( -\frac{1}{2}
\Delta +V_{ext}+v_{H}+v_{xc}\left[ n\right] \right) \phi _{i}=\varepsilon
_{i}\phi _{i},
\end{equation}
where
\begin{equation}
n\left( \vec{x}\right) =\sum_{i}\left( 1+e^{\beta \left( \varepsilon
_{i}-\mu \right) }\right) ^{-1}\left| \phi _{i}\left( \vec{x}\right) \right|
^{2}.
\end{equation}
The chemical potential is fixed by the condition:
\begin{equation}
N=\sum_{i}\left( 1+e^{\beta \left( \varepsilon _{i}-\mu \right) }\right)
^{-1},
\end{equation}
$N$ being the number of conduction electrons. The term $V_{ext}$ represents a uniform background 
potential inside the shell. Its value is fixed such that the converged calculations lead to a 
ionization potential equal to 5.4 eV, appropriate for gold. To avoid problems associated with the 
large degeneracies in spherically symmetric systems, we perform the calculations at low but 
finite temperature. The self-consistency is achieved by employing the usual iterative 
method. In the jellium approximation, the system becomes spherically symmetric and 
consequently the Kohn-Sham orbitals must be of the form 
$r^{-1}u_{lk}\left( r\right) Y_{lm}\left( \hat{r}\right) $.
In this case the Kohn-Sham equations are reduced to a set of coupled radial Schroedinger
equations:
\begin{equation}
\left( -\frac{1}{2} \frac{d^{2}}{dr^{2}}
+\frac{l(l+1) }{2r^{2}}
+v_{eff}[n] \right) u_{lk}(r) =\varepsilon_{lk} u_{lk}(r),
\end{equation}
where
\begin{equation}
n(r) =\frac{2s+1}{r^{2}} \sum_{l,k}\frac{2l+1}{1+e^{\beta \left(
\varepsilon _{lk}-\mu \right) }}\left| u_{lk}\left( r\right) \right| ^{2}.
\end{equation}
Even with the spherical symmetry, in order to make contact with the experiment, one still needs
to calculate a very large number of orbitals. For example, in our largest simulation performed 
so far, the angular quantum number $l$ was as high as 300 and the radial quantum number $k$ of 
the discrete states was as high as 18. The dimensions of this nanoshell were (16,19) nm and,
from an experimental point of view, this will be considered a very small nanoshell. 
The success of such large calculations clearly depends on how fast each iteration is performed. 

\begin{figure}[h]
\begin{center}
\begin{tabular}{c}
\includegraphics[height=12.5cm]{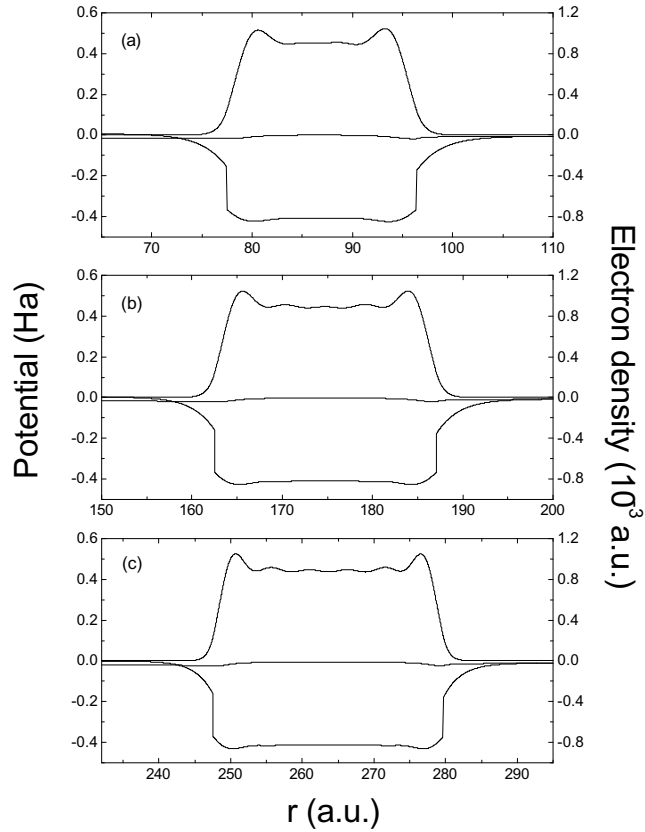}
\end{tabular}
\end{center}  
\caption[example]
{The electronic structure of three different nanoshells which were experimentally fabricated and
characterized by optical measurements. In each panel, the curves represent, from bottom and up, 
the effective and Coulomb potential (left axis) and the electron density (right axis). The dimensions 
of the nanoshells are: (4.1,5.1) nm (panel a), (8.6,9.9) nm (panel b) and (13.1,14.9) nm (panel c).
}
\end{figure}

The radial Kohn-Sham orbitals and 
energies are calculated using the shooting method. This method consists of finding two, 
generally independent, solutions of the radial Kohn-Sham equation corresponding to an
arbitrary energy. One solution is regular in the origin and the other is regular 
at infinity.\cite{ProdanNordlander02CPL} These solutions are found by direct integration 
of the radial Kohn-Sham equation
using the fourth order Runge-Kutta method. The eigenvalues correspond to those particular 
values of the energy where the two solution become linearly dependent, i.e. their Wronskian 
is zero. We notice that this approach provides the eigenvalues and the orbitals in the same time. 
To find all eigenvalues, one will have to sweep the energy from the bottom of the 
effective potential to zero and record the points where the Wronskian changes sign. If there
were no relation between different eigenvalues this will represent a prohibitively time consuming
step. Fortunately, for a shell geometry, the radial Schrodinger equations corresponding to
consecutive angular quantum numbers $l$ have very close eigenvalues and $\varepsilon_{l+1,k}$ is 
always above $\varepsilon_{l,k}$. This means we can reduce the sweeping interval to small intervals
around $\varepsilon_{l,k}$, $k=0,1,2,...$, in order to find the eigenvalues corresponding to $l+1$. 
Thus, all the $l>0$ orbitals can be calculated very efficiently. This is the reason why the shell 
geometry allows calculations with such large number of particles. Suppose we keep the thickness of 
the shell $d$ constant and we increase the outer radius $b$ of the shell. In this case the maximum 
value of the radial quantum number $k$ remains constant while the number of particles increases 
approximately as $N=3db^{2}/r^{3}_{s}$. This means that, as the overall size of the shell is 
increased, only orbitals with higher angular quantum number appear and, as we argued above, they can 
be calculated very efficiently.

Another difficult problem we encountered in our large simulations is the convergence of the
iterative process. It was already pointed out in the literature that the
iterative methods in general have
convergence problem for large many-body systems.\cite{Bertsch90CPC} 
Our studies on the thermodynamic limit of the Kohn-Sham and Hartree equations revealed that this 
convergence problem is due to the long range of the Coulomb 
interaction.\cite{ProdanNordlanderJMPI,ProdanNordlanderJMPII,ProdanNordlander03JSP}
The solution to this problem is to replace the Coulomb 
interaction with a screened Coulomb interaction in the Hartree potential:     
\begin{equation}
v_{H}\rightarrow \int \frac{e^{-\left| \vec{x}-\vec{y}\right| /\lambda }}{%
\left| \vec{x}-\vec{y}\right| }[n\left( \vec{y}\right)-n_{0}] d\vec{y}\text{.}
\end{equation}
The iteration converges fast for a small screening lengths $\lambda$ and suppose the result is 
$n_{\lambda }$. The next step is to increase the screening length $\lambda $ by a finite  
amount and start again the iteration process with $n_{\lambda }$ as the
initial input. This is repeated until the results stabilizes. In the present calculations, the 
screening length was increased by $2$ a.u. and, for each $\lambda $, 8 iterations were needed to 
achieve the convergence. The first four decimals of the eigenvalues become independent
of $\lambda $ for $\lambda \gtrsim 40$ a.u..

Experimentally, the metallic shells are grown over a solid dielectric, which so far was
either Au$_{2}$S 
or silica, and they are usually suspended in solution during measurements. Also, the gold ion cores 
are highly polarizable due to the completely filled d-band. Consequently, the conduction electrons
are embedded in a highly polarizable medium. The effect of these dielectric media can be included 
in the electronic structure calculations as follows. Let us consider a nonuniform, possibly frequency 
dependent dielectric,
\begin{equation}
\varepsilon \left( r\right) =\left\{
\begin{array}{l}
\varepsilon _{1}=\varepsilon _{C}\ \ \text{for }\ r<r_{1} \\
\varepsilon _{2}=\varepsilon _{J}\ \ \text{for}\ \ r_{1}<r<r_{2} \\
\varepsilon _{3}=\varepsilon _{E}\ \ \text{for }\ r_{2}<r,
\end{array}
\right.
\label{eq:dielectric}
\end{equation}
where $r_{1}$ and $r_{2}$ represents the inner, respectively outer radius of the shell.
This parameterization accounts for the presence of a dielectric core,
polarizable jellium background (metal cores) and
a dielectric embedding medium. The Hartree term, 
\begin{equation}
v_{H}(\vec{r})=\int
\frac{n(\vec{r}\,^{\prime })-n_{0}}{| \vec{r}-\vec{r}\,^{\prime }| }
d\vec{r}\,^{\prime },
\end{equation}
which essentially represents the electrostatic potential generated by
the conduction electrons and the 
positive ion cores, has to be modified such that it satisfies the usual boundary conditions at the 
interface between different dielectric media:\cite{RubioSerra93PRB}
\begin{equation}
\tilde{v}_{H}(\vec{r})=\left\{
\begin{array}{l}
v_{H}(\vec{r})/\varepsilon _{C}+\Phi _{C}\text{ \ \ for \ \ }\left| \vec{r}\right| <r_{1} \\
v_{H}(\vec{r})/\varepsilon _{J}+\Phi _{J}\text{ \ \ for \ \ }r_{1}<\left| \vec{r}\right|
<r_{2} \\  
v_{H}(\vec{r})/\varepsilon _{E}\text{ \ \ for \ \ }r_{2}<\left| \vec{r}\right| ,
\end{array}   
\right.
\end{equation}
The constants are found from the requirement that the potential is continuous and they are given 
by:
\begin{equation}   
\left\{
\begin{array}{l}
\Phi _{J}=\left(   
1/\varepsilon _{E}-
1/\varepsilon _{J}
\right) v_{H}\left( r_{2}\right)  \\
\Phi _{C}=\Phi _{J}+\left(
1/\varepsilon _{J}-
1/\varepsilon _{C}
\right) v_{H}\left( r_{1}\right) \text{.}
\end{array}
\right.
\end{equation}
In Fig.~1, we show the effect of the dielectric media on the electronic
structure of the nanoshell. The values for the dielectric constants were chosen as 
$\varepsilon_{C}=5$, $\varepsilon_{E}=2$ and $\varepsilon_{J}=8$ which are representative
for realistic situations. The influence of $\varepsilon_C$ and $\varepsilon_E$
is relatively weak. The strongest influence on the electronic structure comes from
$\varepsilon_J$ which strongly reduces the dipole fields near the surfaces. This leads to a lowering 
of the effective potential in the nanoshell. The reduced workfunction results in an
increased electron spill out at the surfaces of the shell. The effect of the d-electrons is 
slightly reduced when the core and embedding medium are added. Although the dielectric effects on the 
electronic structure are visible, these variations alone have little influence on the optical 
response of the nanoparticles.\cite{ProdanetAl02CPL} This is in contrast to what have been observed 
experimentally, 
i.e. large shifts in the plasmon frequencies due to the presence of different dielectric media. As 
will be argued in the next section, the explanation is that the system responds differently to an 
external electric field in the presence of the dielectrics which screen the excitation.

Figure~2 illustrates the converged electronic structure of three nanoshells which were 
fabricated and characterized   
by optical measurements. The electronic structure shown in this figure will be later used to 
calculate the optical response of these nanoshells and make a direct comparison with the 
experiment. The dimensions of these nanoshells are (4.1,5.1) nm, (8.6,9.9) nm and (13.1,14.9) 
nm. The largest nanoparticle contains about $2.5 \times 10^{5}$ conduction electrons. The nanoshells 
were suspended in an aqueous solution ($\varepsilon_{E}=1.78$) and their core
consisted of Au$_{2}$S ($\varepsilon_{C}=5.4$). The value of the jellium dielectric constant was 
fixed to $\varepsilon_{J}=8$, determined from a fit of the experimentally measured gold dielectric 
function with a Drude like expression.

The independent electron response function is calculated directly from the electronic structure. 
We are interested here only in the dipolar term ($l=1$) of the response function, which is calculated 
using the following expansion:\cite{ZangwillSoven80PRA}
\begin{eqnarray}
\tilde{\Pi}_{l=1}^{0}\left( r,r^{\prime };\omega \right)=
\sum_{|l_{1}-l_{2}|=1}\frac{l_{1}+l_{2}+1}{4\pi }
\sum_{k}f_{\beta }\left( \varepsilon _{kl_{1}}\right)
u_{kl_{1}}\left( r\right) u_{kl_{1}}\left( r^{\prime }\right) \nonumber \\
\times \left[ G_{l_{2}}\left( r,r^{\prime };\varepsilon _{kl_{1}}+\omega
+i\delta \right) +G_{l_{2}}\left( r^{\prime },r;\varepsilon
_{kl_{1}}-\omega
-i\delta \right) \right] \text{,}
\label{eq:pi0tilde}
\end{eqnarray}
where $G_{l}$, $l=0,1,...,$ represent the radial Green functions:
\begin{equation}
\left( -\frac{1}{2}\frac{d^{2}}{dr^{2}}+\frac{l\left( l+1\right) }{2r^{2}}
+v_{eff}\left( r\right) -E-i\delta \right) G_{l}\left( r,r^{\prime };E+i\delta \right)
=-\delta \left( r-r^{\prime }\right) \text{,}  \label{GreenEq}
\end{equation}
and $u_{kl}$ are the self-consistently calculated Kohn-Sham orbitals. The Green functions are 
calculated by direct integration of their specific equation. Eq. (\ref{eq:pi0tilde}) provides 
the response function multiplied by $r^{2}r^{\prime 2}$ which is more convenient to use as it will 
follow from the next section.

\section{Optical response}

We will focus in this section on deriving the RPA equations in the presence of the dielectric 
medium defined by Eq. (\ref{eq:dielectric}). The physical picture can be described as follow. When 
the whole system is placed in an external field, the conduction electrons will try to screen the 
excitation field. The dielectrics will polarize when placed in the external field, screening at their 
turn the excitation field. The conduction electrons will screen not only the original excitation but 
also the electric fields produced by the polarized dielectrics. To quantify these processes, we start 
from the TDLDA expression of the screening charge induced by an external electric field
$ E_{0}e^{-i\omega t}\,\vec{e}_{z}$,
\begin{equation}
\delta n(\vec{r},\omega )=\int d\vec{r^{\prime }}\Pi ^{(0)}(\vec{r},\vec{r}%
\,^{\prime };\omega )\left[ \delta v_{xc}\left( \vec{r}\,^{\prime },\omega
\right) +\phi _{C}\left( \vec{r}\,^{\prime },\omega \right) \right] ,
\label{TDLDA}
\end{equation}
where $\Pi ^{\left( 0\right) }$ represents the response function of the
independent electrons. The quantity $\delta v_{xc}$ represents the variation
of the exchange-correlation potential due to the screening charge and $\phi _{C}$
represents the Coulomb potential generated by the screening charges $\delta n$ and including the
excitation field. In the quasistatic limit, which is generally accepted to be valid
for particles smaller than 40~nm,\cite{KreibigVollmer95Book,Sonnichsen02NJP} the
electrtatic potential $\phi _{C}$ must satisfy the Poisson equation with the appropriate
boundary conditions:
\begin{equation}
\left\{
\begin{array}{l}
\varepsilon _{i}\vec{\nabla}^{2}\phi _{C}\left( \vec{r},\omega \right)
=-4\pi \delta n\left( \vec{r},\omega \right)  \\
\varepsilon _{i}\left. {{\frac{\partial \phi _{C}}{\partial \vec{n}}}}
\right| _{r=r_{i}^{-}}=\varepsilon _{i+1}\left. {{\frac{\partial \phi _{C}}
{ \partial \vec{n}}}}\right| _{r=r_{i}^{+}} \\
\phi _{C}\left( \vec{r},\omega \right) \rightarrow -\vec{r}\vec{E}_{0}
\ for \ \left| \vec{r}\right| \rightarrow \infty .
\end{array}
\right.   \label{Poisson}
\end{equation}
The solution can be written in the following form:
\begin{equation}
\phi _{C}\left( \vec{r},\omega \right) =\left( -r+\frac{4\pi }{3}\int \frac{r_{<}}
{r_{>}^2 \varepsilon \left( r^{\prime }\right) }\delta n\left( r^{\prime
},\omega \right) r^{\prime 2}dr^{\prime }
+\sum\nolimits_{i=1,2}r_{i}^{2}\sigma ^{\left( i\right) }v_{1}\left(
r,r_{i}\right) \right) E_{0}\cos \theta \text{.}  \label{Potential}
\end{equation}
The surface charges $\sigma^{1,2}$ are found from the boundary condition as described in a previous 
application.\cite{ProdanetAl03CPL} After the solution is plugged back into Eq. (\ref{TDLDA}) and 
using the notation $v_{1}\left( r,r^{\prime }\right) =r_{<}/r_{>}^{2}$,
$v_{2}\left( r,r^{\prime}\right) =dv_{1}/dr$ and $\alpha \left( r,\omega \right) =r^{2}\delta
n\left( r,\omega \right) $, the RPA equation becomes:
\begin{eqnarray}
&&\alpha \left( r,\omega \right) -\int dr^{\prime }\,\tilde{\Pi}_{1}^{\left(
0\right) }\left( r,r^{\prime };\omega \right) r^{\prime -2}v_{xc}^{\prime
}\left( n\left( r^{\prime }\right) \right) \alpha \left( r^{\prime },\omega
\right)  \\
&&-\frac{4\pi }{3}\int dr^{\prime }\int dr^{\prime \prime }\,\tilde{\Pi}%
_{1}^{\left( 0\right) }\left( r,r^{\prime };\omega \right) \frac{v_{1}\left(
r^{\prime },r^{\prime \prime }\right) }{\varepsilon \left( r^{\prime \prime
}\right) }\alpha \left( r^{\prime \prime },\omega \right)   \nonumber \\
&&-\frac{4\pi }{3}\sum_{i=1,2}X^{\left( i\right) }\left( r,\omega \right)
\int \frac{v_{2}\left( r_{i},r^{\prime }\right) }{\varepsilon \left(
r^{\prime }\right) }\alpha \left( r^{\prime },\omega \right) dr^{\prime }
\nonumber \\
&=&\int dr^{\prime }\,\tilde{\Pi}_{1}^{\left( 0\right) }\left( r,r^{\prime
};\omega \right) \phi _{C}^{0}\left( r^{\prime }\right) ,  \nonumber
\label{eq:pol}
\end{eqnarray}
where
\begin{eqnarray}
X^{\left( 1\right) }\left( r,\omega \right)  &=&-\frac{1}{\xi _{\varepsilon }}
\frac{\varepsilon _{C}-\varepsilon _{J}}{\varepsilon _{C}+2\varepsilon _{J}}
X_{1}\left( r,\omega \right) -\frac{\xi _{\varepsilon }-1}{\xi _{\varepsilon
}}X_{2}\left( r,\omega \right)  \\
X^{\left( 2\right) }\left( r,\omega \right)  &=&\frac{\xi _{\varepsilon }-1}
{ 2\xi _{\varepsilon }}\left( \frac{r_{2}}{r_{1}}\right) ^{3}X_{1}\left(
r,\omega \right) +\frac{1}{\xi _{\varepsilon }}\frac{\varepsilon
_{E}-\varepsilon _{J}}{\varepsilon _{J}+2\varepsilon _{E}}X_{2}\left(
r,\omega \right) .  \nonumber
\end{eqnarray}
and
\begin{equation}
X_{i}\left( r,\omega \right) =r_{i}^{2}\int dr^{\prime }\,\tilde{\Pi}%
_{1}^{\left( 0\right) }\left( r,r^{\prime };\omega \right) v_{1}\left(
r^{\prime },r_{i}\right) .
\end{equation}
In these equations, $\phi _{C}^{0}$ represents the unscreened response of
the dielectrics to the electric field:
\begin{equation}
\phi _{C}^{0}\left( \vec{r}\right) =\left(
-r+\sum\limits_{i=1,2}r_{i}^{2}\sigma _{0}^{\left( i\right) }v_{1}\left(
r,r_{i}\right) \right) E_{0}\cos \theta,
\end{equation}
where
\begin{eqnarray}
\sigma _{0}^{\left( 1\right) } &=&\frac{1}{\xi _{\varepsilon }}\frac{%
\varepsilon _{C}-\varepsilon _{J}}{\varepsilon _{C}+2\varepsilon _{J}}\frac{%
3\varepsilon _{E}}{\varepsilon _{J}+2\varepsilon _{E}} \\
\sigma _{0}^{\left( 2\right) } &=&1-\frac{1}{\xi _{\varepsilon }}\frac{%
3\varepsilon _{E}}{\varepsilon_{J}+2\varepsilon _{E}}  \nonumber
\end{eqnarray}
are the classical polarization charges and $\xi _{\varepsilon}$ is just a
dielectric factor,
\begin{equation}
\xi _{\varepsilon }=1-2\frac{\varepsilon _{C}-\varepsilon _{J}}{\varepsilon
_{C}+2\varepsilon _{J}}\frac{\varepsilon _{E}-\varepsilon _{J}}{\varepsilon
_{J}+2\varepsilon _{E}}\left( \frac{r_{1}}{r_{2}}\right) ^{3}.
\end{equation}
The integrals in the above equations are discretized and the integral equations are 
transformed in algebraic equations which are solved by conventional  
methods.\cite{ProdanNordlander02CPL,ProdanetAl02CPL} The photo absorption cross section of the 
nanoshell is related to its polarizability $\alpha (\omega )$ through 
\begin{equation}
\sigma_{abs}(\omega )=\sqrt{\varepsilon_{E}} \dfrac{\omega}{c} Im[\alpha (\omega )],
\end{equation}
where SI units were used. The frequency dependent polarizability is calculated from
\begin{equation}
\alpha \left( \omega \right) =\frac{4\pi }{3}\int dr\,r^{3}\alpha \left(
r,\omega \right) .  \label{polariz}
\end{equation}
where $\alpha \left( r,\omega \right) $ is the local polarizability derived from the RPA 
equation Eq. (\ref{eq:pol}). The absorbance is defined by the ratio between the optical absorption 
cross section and $\pi r_{2}^{2}$.

\begin{figure}
\begin{center}
\begin{tabular}{c}
\includegraphics[height=8cm]{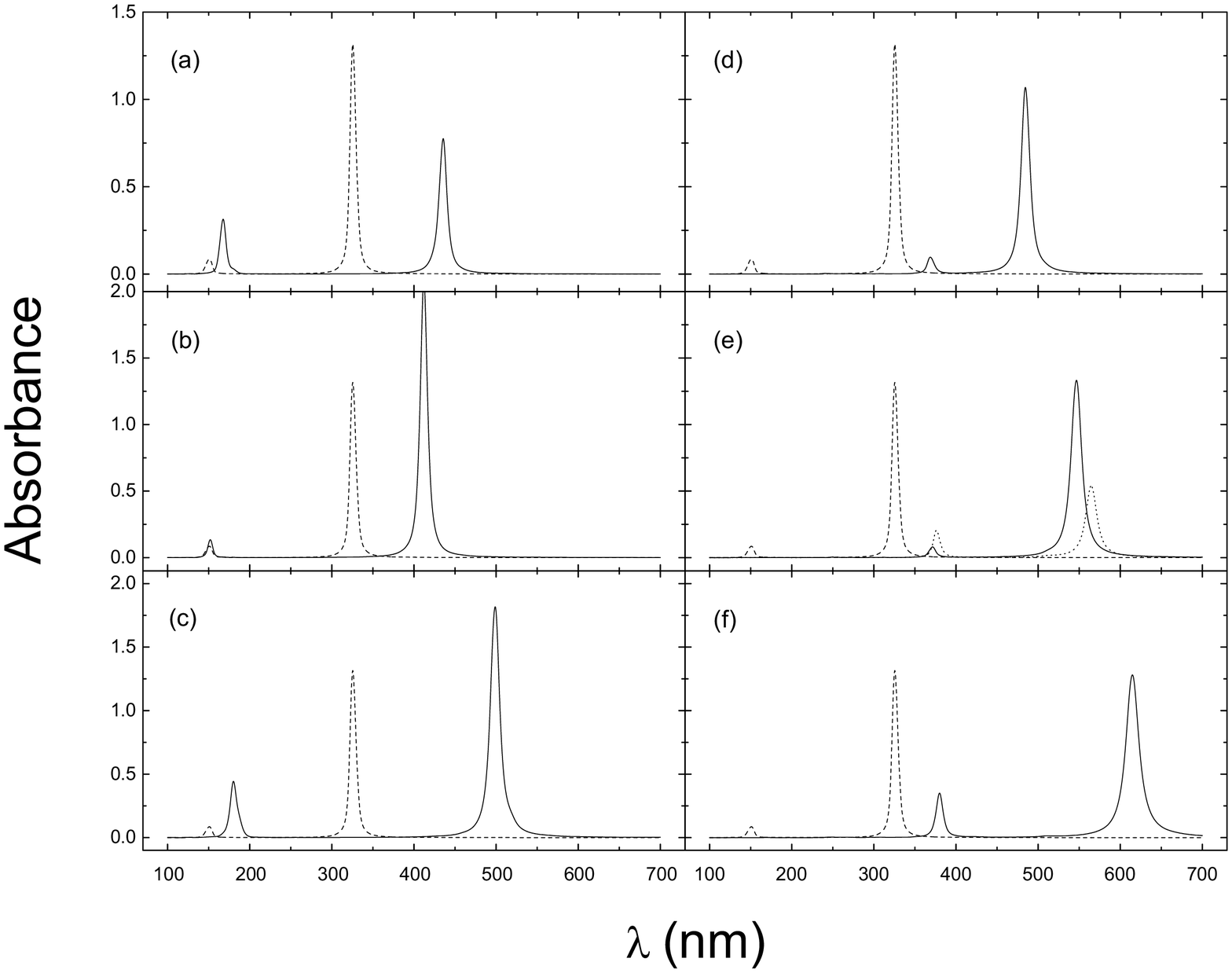}
\end{tabular}
\end{center}
\caption[example]
{The effect of different dielectrics on the optical absorption of the gold nanoshell
shown in Fig. 1.  Different panels refer to the following situations:
a) $\varepsilon_{C}=5$, $\varepsilon_{E}=1$ and
$\varepsilon_{J}=1$; b) $\varepsilon_{C}=1$, $\varepsilon_{E}=2$ and
$\varepsilon_{J}=1$; c)
$\varepsilon_{C}=5$, $\varepsilon_{E}=2$ and $\varepsilon_{J}=1$;
d) $\varepsilon_{C}=1$, $\varepsilon_{E}=1$
and $\varepsilon_{J}=8$; e) $\varepsilon_{C}=1$, $\varepsilon_{E}=2$ and
$\varepsilon_{J}=8$ (solid) and
$\varepsilon_{C}=5$, $\varepsilon_{E}=1$ and $\varepsilon_{J}=8$ (dotted);           
f) $\varepsilon_{C}=5$, $\varepsilon_{E}=2$ and $\varepsilon_{J}=8$.
For comparison, in all panels we have included the absorbance of
the gold nanoshell when no core or embedding medium is present and
$\varepsilon_{J}=1$ (dashed line).
}
\end{figure} 

\begin{figure}
\begin{center}
\begin{tabular}{c}
\includegraphics[height=8cm]{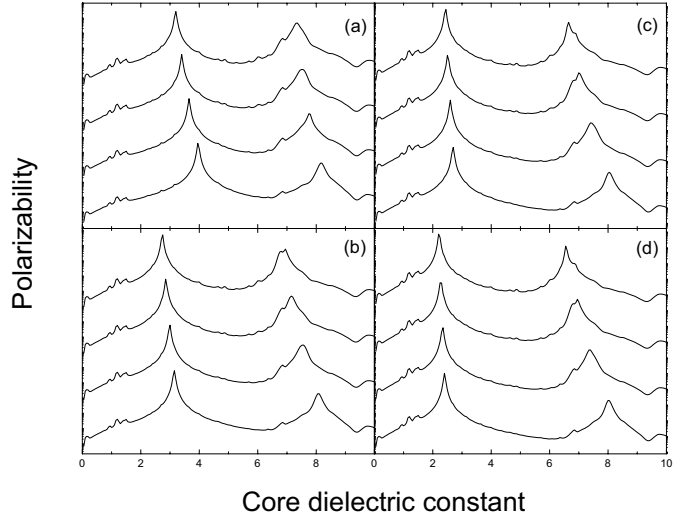}
\end{tabular}
\end{center}
\caption[example]
{ \label{fig3}
The optical response of a (60,90) a.u. gold nanoshell when the dielectric constant
of the core and embedding medium are varied over a range of 1 to 4. The dielectric
constant of the jellium shell $\varepsilon_J=1$. A logarithmic scale is used on the
y=axis. Each panel,
from the bottom curve and up, represents the optical response corresponding to
a dielectric core of 1,2,3 and 4. Different panels corresponds to
$\varepsilon_{E}=1$ ( panel a), $\varepsilon_{E}=2$ (panel b),
$\varepsilon_{E}=3$ (panel c) and $\varepsilon_{E}=4$ (panel d).
}
\end{figure}

\begin{figure}
\begin{center}
\begin{tabular}{c}
\includegraphics[height=10cm]{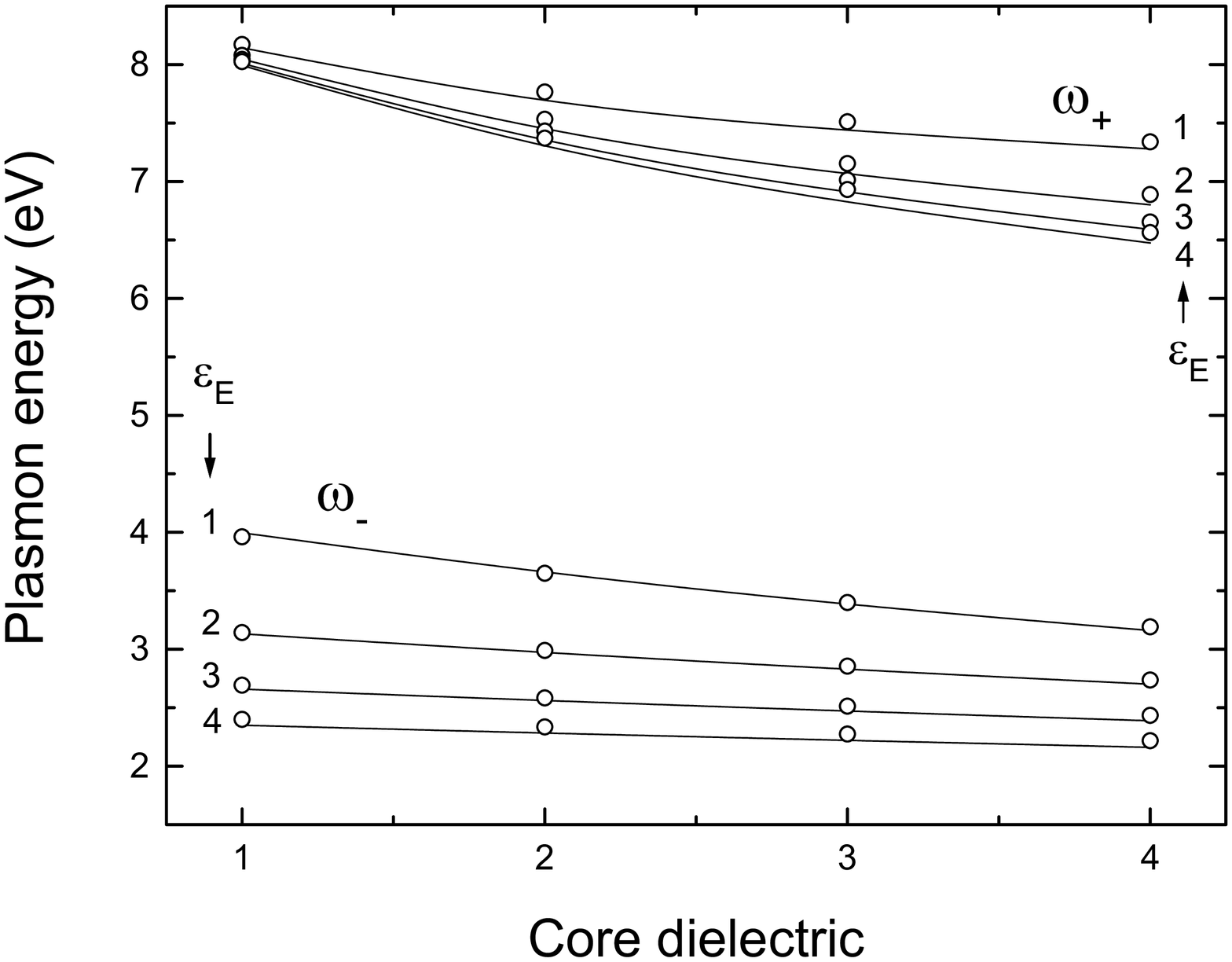}
\end{tabular}
\end{center}
\caption[example]
{
The TDLDA calculated plasmon energies (circles) compared with the prediction of
the classical Mie theory (solid lines) for different core dielectrics and embedding media.
}
\end{figure}

In Fig.~3, we show how the photoabsorption cross section of a (70,100)~a.u. nanoshell
is influenced  by the presence of dielectric media.  For each nanoshell, the
photoabsorption spectrum is dominated by two peaks, corresponding to the symmetric
and antisymmetric coupling between charges on the inner and outer surfaces of the shell. The
nanoshell plasmons can be viewed as bonding and antibonding superpositions of a cavity plasmon
located on the inner surface of the shell and a sphere surface plasmon on the outer surface of the
shell. In the presence of dielectric backgrounds, the energy of the cavity plasmon is
$\omega_{C} =\omega_B\sqrt{\frac{2}{\varepsilon_C + 2\varepsilon_J}}$
and the energy of the sphere plasmon is
$\omega_{S}=\omega_B\sqrt{\frac{1}{2\varepsilon_E +\varepsilon_J}}$
The low energy (long wavelength) symmetric nanoshell plasmon $\omega_-$
has a larger admixture of the sphere plasmon and the high energy (low wavelength) antisymmetric mode
$\omega_+$ is primarily composed of the the cavity plasmon. The changes in the optical spectra in
Fig.~3  caused by the different dielectric media can be simply understood from how the bare cavity
and sphere plasmon shifts in the presence of dielectrics. The $\omega_+$ plasmon with its large
cavity plasmon content is more strongly influenced by a dielectric core $\varepsilon_C$ than by the
embedding medium. The $\omega_-$ plasmon is more strongly influenced by the dielectric embedding
medium $\varepsilon_E$ than by the dielectric core. The presence of a dielectric jellium background
$\varepsilon_J$ has a stronger influence on $\omega_+$ simply because the cavity plasmon
depend more strongly on $\varepsilon_J$ than the surface sphere plasmon.

In Fig.~4 we calculate the optical absorption spectra of a (60,90) a.u. for a broad range of
values of the core and embedding medium dielectric constant. The optical absorption is plotted in a
logarithmic scale. The figure show that in addition to the two collective resonances $\omega_+$ and
$\omega_-$ there is a background contribution to the optical absorption. This background
absorption is caused by discrete single particle excitations. It can be seen that as the dielectric
constant of the core and embedding medium is changed, only the collective plasmon resonances are
influenced. The discrete single particle excitations remain essentially unchanged because the
presence of dielectric media only weakly influences the electronic structure of the system.
A calculation of the polarizability only including the changes in the electronic
structure caused by the dielectrics has almost no effect on the optical properties of
the nanoshell.\cite{ProdanetAl02CPL} The large effect of a dielectric core or
embedding medium on the plasmon resonances is due to the polarization of the dielectrics and their
contribution to $\phi_C$ in Eq.~(\ref{TDLDA}).

The frequency of the plasmon resonances extracted from the optical absorption spectra shown in 
figure Fig.~4 are in excellent agreement with the predictions of
the classical Mie theory as it is 
illustrated in Fig.~5. In the classical calculations, the metallic phase is modeled by a pure Drude 
dielectric function corresponding to the same $r_{s}=3$ as used in the TDLDA calculations. 

\begin{figure}
\begin{center}
\begin{tabular}{c}
\includegraphics[height=10cm]{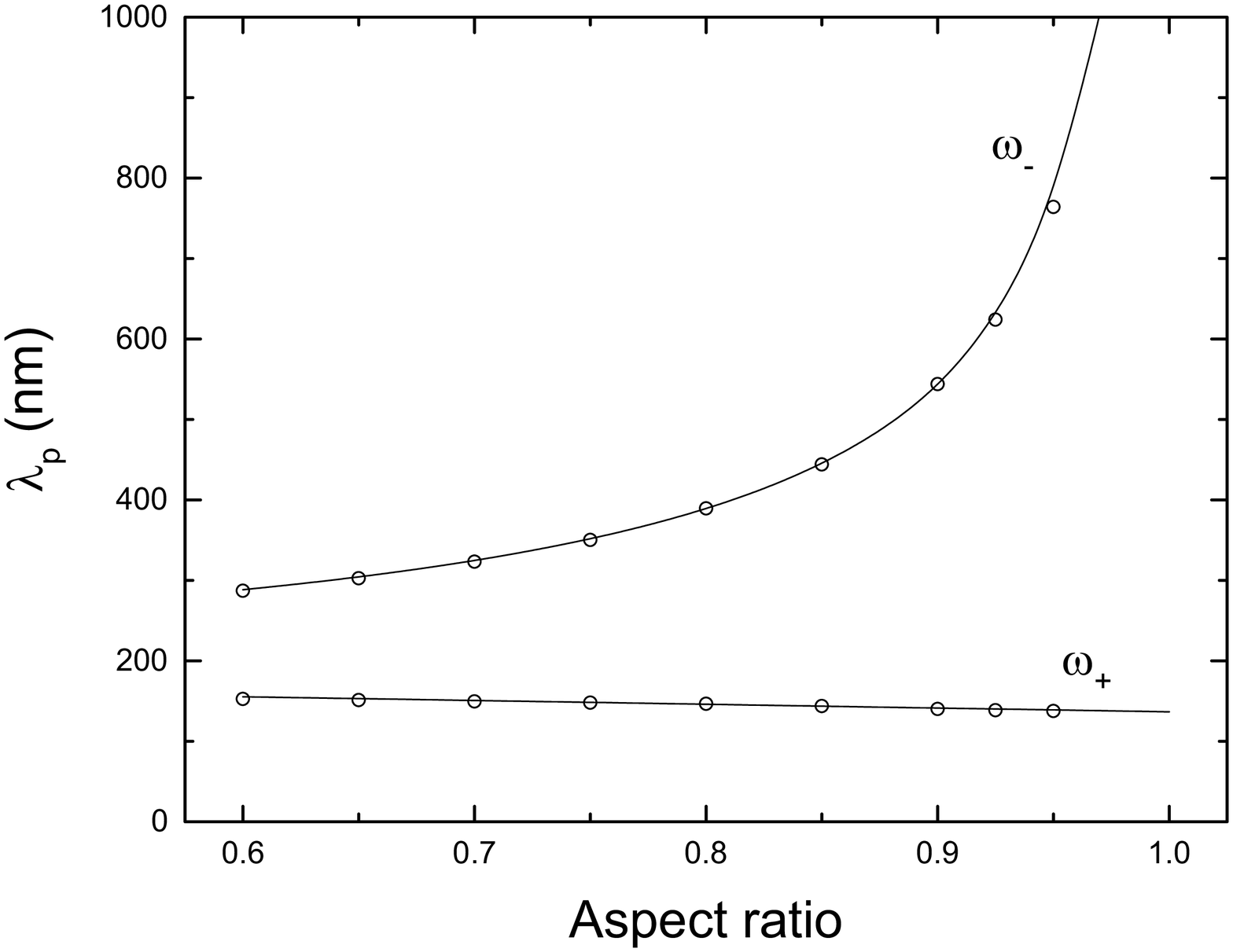}
\end{tabular}
\end{center}
\caption[example]
{ \label{fig4}
The TDLDA calculated plasmon wavelengths (circles) compared with the prediction of
the classical Mie theory Eq. (\ref{eq:sca}) (solid lines) for nanoshells of different aspect 
ratios. The TDLDA has been applied to a set of nine nanoshells of the same thickness (17 a.u.) and 
with overall diameters of 85, 97.2, 117.4, 136, 170, 226.6, 340, 453 and 680 a.u..
}
\end{figure}

The fundamental property of the metallic nanoshells is their tunable plasmon resonances, which was 
demonstrated in a series of experiments where monodisperse nanoshells of different 
shell thickness and core sizes have been fabricated and characterized with optical
measurements.\cite{OldenburgetAl99APL} In the absence of any dielectrics, the classical Mie theory 
predicts the following expression for the plasmon energies:
\begin{equation}
\omega _{l\pm }^{2}=\dfrac{\omega _{B}^{2}}{2}[1\pm \frac{1}{2l+1}
\sqrt{1+4l(l+1)x^{2l+1}}],  
\label{eq:sca}
\end{equation}
which depends only on the aspect ratio of the nanoshell $x=r_{1}/r_{2}$. The same expression can be 
derived using semi-classical approaches.\cite{MukhopadhyayLundqvist75INC} We show in 
the following that our self-consistent calculations lead to the same aspect ratio dependence of 
the plasmon energies. For this, we fix the dielectric constants to 1 and we apply the TDLDA formalism 
to nine nanoshells of the same thickness (17 a.u.) but with different overall diameters, ranging from 
85 to 680 a.u.. The aspect ratio of the nanoshell is varied in this way over a broad range of 
values starting from 0.6 and up to 0.95. From the polarizability curves we extract the wavelengths of 
the collective modes which are shown in Fig.~6 by open circles. The agreement between the TDLDA 
calculated plasmon wavelengths and Eq. (\ref{eq:sca}) is almost perfect over the entire range of 
aspect ratios. The lower energy plasmon becomes strongly dependent on the aspect ratio for 
thin shells when the interaction between the cavity and sphere plasmon becomes very strong. This is 
the regime where the nanoshells become very useful in a variety of applications.

\section{Comparison with experimental data}

\begin{figure}[h]
\begin{center}   
\begin{tabular}{c}
\includegraphics[height=12cm]{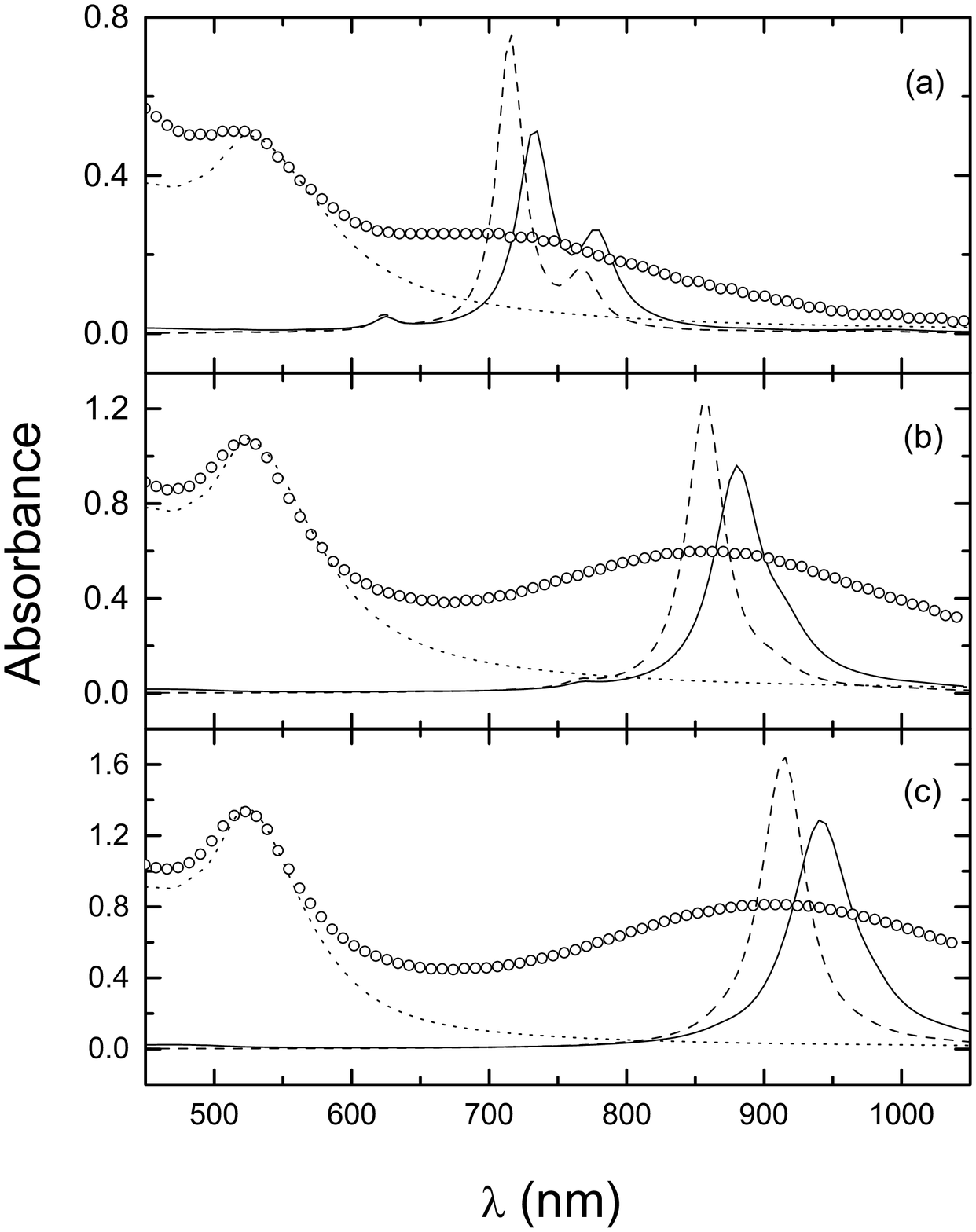}
\end{tabular}
\end{center} 
\caption[example]
{ \label{fig5}   
The optical response of three gold nanoshells with Au$_{2}S$ core of geometries (4.1,5.1)~nm
(panel a), (8.6,9.9)~nm (panel b) and (13.1,14.9)~nm (panel c).
The solid curves represent the TDLDA calculation when the
full frequency dependent dielectric constant of gold is used. The dashed lines
represent the TDLDA calculation when a Drude dielectric function with
$\varepsilon_{J}=8$ is used. The circles represent the experimentally
measured optical response. The dotted lines represents the optical absorption of
solid gold particles also
present in the solution during the experiment.

}
\end{figure}

In this section we apply the formalism presented in the last section and calculate the optical 
response of the three nanoshells presented in Fig.~2. The procedure for calculating an effective
jellium background
dielectric function consists of the following. The conduction electrons are assumed to contribute 
to the dielectric function of the metal through a Drude term:
\begin{equation}
\varepsilon_{D}(\omega)=1-\omega_{B}^{2}/(\omega^2+i\gamma \omega).
\end{equation}
This assumption is supported by the excellent agreement between the TDLDA calculations and the 
classical Mie calculations when the metallic phase is modeled by a Drude dielectric function. The 
width $\gamma$ should be the same as the artificial width that will be used in the TDLDA
calculations.\cite{LermeetAl99PRB} The experimentally measured bulk dielectric 
function of gold,\cite{JohnsonChristy72PRB} $\varepsilon^{exp}(\omega)$ can then be decomposed 
into:\cite{LermeetAl99PRB,ShahbazyanAtal98PRL}
\begin{equation}
\varepsilon^{exp}(\omega)=\varepsilon_{J}(\omega)+\varepsilon_{D}(\omega)-1,
\end{equation}
which provides a simple way of determining the jellium dielectric
function $\varepsilon_{J}(\omega)$.
The RPA equation Eq. (\ref{eq:pol}) have been solved with the frequency
dependent jellium dielectric 
function obtained by this procedure and the results are shown in
Fig.~7 by solid lines. We also show a calculation using a frequency independent
$\varepsilon_J=8$ obtained from a best fit of $\varepsilon_J(\omega)$ to a constant.
The agreement 
with the experimentally measured optical absorption shown by circles is excellent for all three 
nanoshells. The experimental spectra show an additional peak for all three nanoshells which is 
located at the same wavelength of about 520~nm. 
This peak is generated by solid gold colloids also present in the solution
during the experiment. The 
peak at larger wavelengths is the $\omega_{-}$ resonant mode and its position is different for the 
three nanoshells, which have different
aspect ratios. The $\omega_{-}$ mode in the TDLDA calculations 
is slightly red shifted relative to the experiment.

\begin{figure}[h]
\begin{center}
\begin{tabular}{c}
\includegraphics[height=12cm]{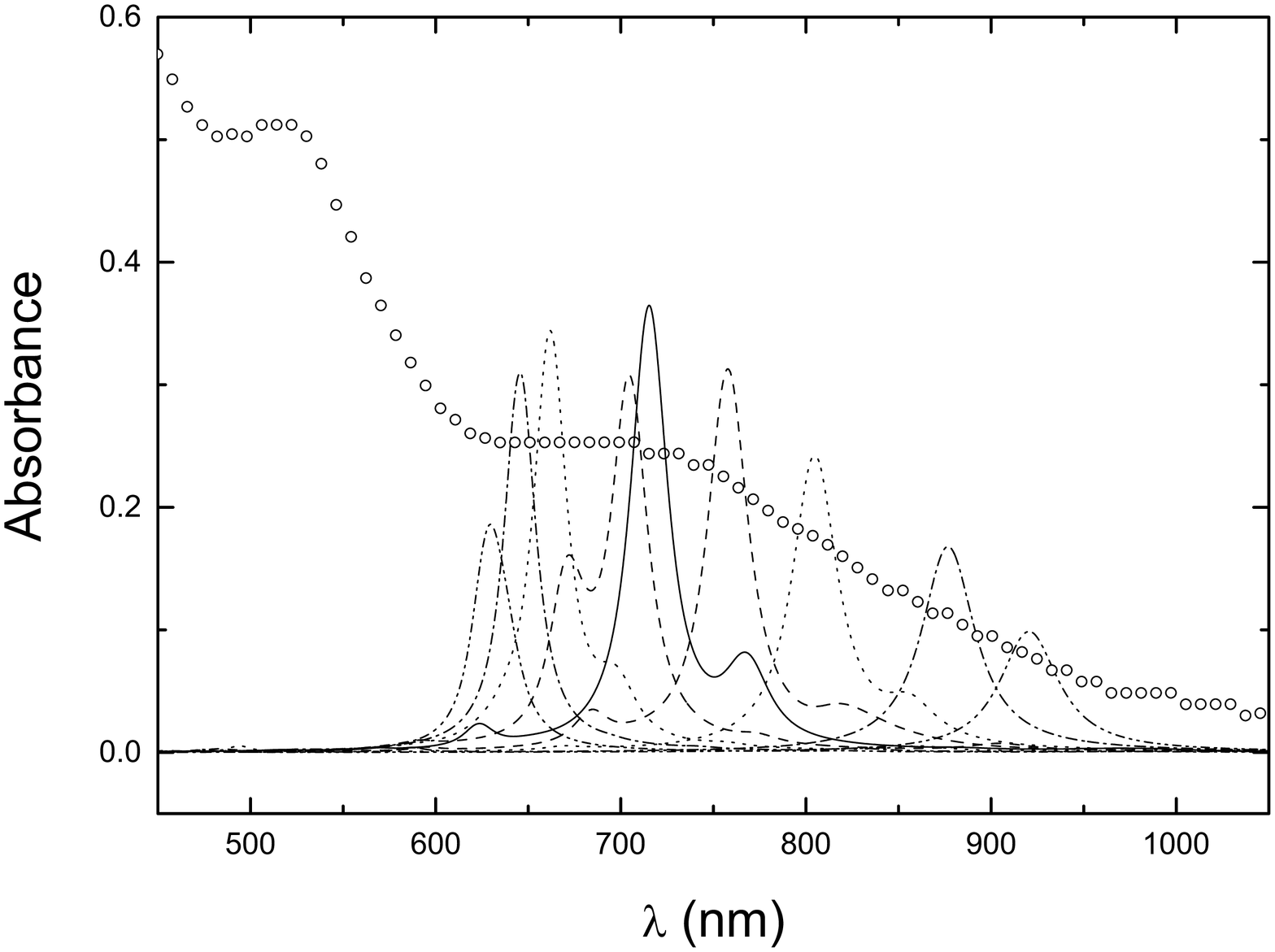}
\end{tabular}
\end{center}
\caption[example]
{ \label{fig6}
The effect of a size distribution on the optical absorption of a solution of gold
nanoshells. The figure illustrates the TDLDA optical absorption of different nanoshells
of slightly different geometries, $\delta=\pm 2.75\%$ (dashed), $\delta=\pm 5.5\%$ (dotted),
$\delta=\pm 8.25\%$ (dashed-dotted) and $\delta=\pm 11\%$ (dashed-dotted-dotted) relative
to the geometry (4.1,5.1) nm (solid). Each curve has been multiplied by a
Gaussian factor $\alpha e^{-(\delta / \sigma)^2}$, where $\sigma=11\%$ as predicted 
from experiment and $\alpha$ was chosen equal to 0.5. The circles represent the
experimentally measured optical absorption.
}
\end{figure}

In these experiments the nanoshells are not perfectly monodisperse and this is due to slight
variations in their geometry. The size distribution was evaluated to a Gaussian of a standard
deviation of $11\%$, which can explain the difference in the widths of the plasmon resonances. To
support this argument, we have calculated the optical
response of nine nanoshells of sizes uniformly
distributed in between $\pm 11\%$ of the standard size.
The results are shown in Fig.~8. The optical
absorption curves have been multiplied by the corresponding Gaussian weight. As one can see, a size
distribution of $11\%$ introduces a large broadening of the plasmons. This is mainly due to the fact
that the thickness of the nanoshells is very small compared with the overall size and, in this
regime, the plasmon modes are very sensitive to small variations of the geometry as it was
illustrated in Fig.~6. Also, the size distribution induces a blue shift because the height of the
plasmon peaks increases with the size of
the nanoshells.
The figure clearly shows that the variation of the plasmon energies caused by
the inhomogeneous size  distribution can account for the experimental broadening of the
plasmon resonance.

\section{Conclusions}

The optical properties of large nanoshells can be calculated using the TDLDA method.
The results show that the plasmon energies are sensitive to geometry and can depend strongly on the 
presence of dielectric backgrounds such as a dielectric core, jellium background and embedding 
medium. The calculated energies of the plasmon resonances are in excellent agreement
with classical Mie scattering. A comparison with experimental data show very good
agreement between the theoretical and experimental absorption spectra.

\acknowledgments
This work was supported by the Robert A. Welch foundation under grant C-1222,
by the Texas Advanced Technology Program,
and by the Multi-University Research Initiative of the Army Research Office.

\end{document}